\documentclass[twocolumn,preprintnumbers,amsmath,amssymb,showpacs]{revtex4}
\usepackage{graphicx}
\usepackage{amsmath}
\usepackage{epsfig}

\begin{document}

\title{Radiative seesaw in left-right symmetric model}

\author{Pei-Hong Gu$^{1}_{}$}
\email{pgu@ictp.it}

\author{Utpal Sarkar$^{2}_{}$}
\email{utpal@prl.res.in}

\affiliation{$^{1}_{}$The Abdus Salam International Centre for
Theoretical Physics, Strada Costiera 11, 34014 Trieste, Italy\\
$^{2}_{}$Physical Research Laboratory, Ahmedabad 380009, India}

\begin{abstract}

There are some radiative origins for the neutrino masses in the
conventional left-right symmetric models with the usual bi-doublet
and triplet Higgs scalars. These radiative contributions could
dominate over the tree-level seesaw and could explain the observed
neutrino masses.

\end{abstract}


\maketitle

\emph{Introduction:} Strong evidence from the neutrino oscillation
experiments has confirmed the tiny but nonzero neutrino masses. This
phenomenon is elegantly explained by the seesaw mechanism
\cite{minkowski1977} in some extensions of the standard model (SM).
The seesaw scenario can be naturally embedded into the left-right
symmetric models \cite{ps1974} and also the grand unified theories
(GUTs).

In this paper, we discuss the neutrino mass generation in a general
class of left-right symmetric models with the Higgs sector including
one bi-doublet, one left-handed triplet and one right-handed triplet
Higgs fields. Our analysis shows that the neutrino masses can
originate from some loop diagrams in addition to the tree-level
seesaw. We also demonstrate that the radiative neutrino masses could
explain the experimental results for some choice of parameters.

\vspace{1mm}

\emph{The left-right symmetric model:} We consider the left-right
symmetric extension of the SM with the gauge group
$SU(2)_{L}^{}\times SU(2)_{R}^{}\times U(1)_{B-L}^{}$ and the
following Higgs content:
\begin{eqnarray}
\phi(\textbf{2},\textbf{2}^{\ast}_{},0)\,,~\Delta_{L}^{}(\textbf{3},\textbf{1},-2)\,,
~\Delta_{R}^{}(\textbf{1},\textbf{3},-2)\,.
\end{eqnarray}
A convenient representation of these fields is given by the $2\times
2$ matrices:
\begin{eqnarray}
\phi = \left[
\begin{array}{cc}
\phi^{0}_{1}&\phi^{+}_{2}\\
\phi^{-}_{1}&\phi^{0}_{2}\end{array} \right]\,,~\Delta_{L,R}^{} =
\left[
\begin{array}{cc}
\delta^{+}_{}/\sqrt{2} &\delta^{++}_{}\\
\delta^{0}_{} &-\delta^{+}_{}/\sqrt{2}\end{array}
\right]_{L,R}^{}\,.
\end{eqnarray}
As for the fermion sector, it includes the left- and right-handed
quarks:
\begin{eqnarray}
q_{L}^{}(\textbf{2},\textbf{1},\frac{1}{3}) = \left[
\begin{array}{c}
u\\
d\end{array} \right]
_{L}^{}\,,~q_{R}^{}(\textbf{1},\textbf{2},\frac{1}{3}) = \left[
\begin{array}{c}
u\\
d\end{array} \right] _{R}^{}\,
\end{eqnarray}
and the left- and right-handed leptons:
\begin{eqnarray}
\psi_{L}^{}(\textbf{2},\textbf{1},-1) = \left[
\begin{array}{c}
\nu\\
\ell\end{array} \right]
_{L}^{}\,,~\psi_{R}^{}(\textbf{1},\textbf{2},-1) = \left[
\begin{array}{c}
\nu\\
\ell\end{array} \right] _{R}^{}\,.
\end{eqnarray}

Under the left-right (parity) symmetry, we have $\phi\leftrightarrow
\phi^\dagger_{}$, $\Delta_{L}^{}\leftrightarrow\Delta_{R}^{}$,
$q_{L}^{}\leftrightarrow q_{R}^{}$ and $\psi_{L}^{}\leftrightarrow
\psi_{R}^{}$. The parity invariant Yukawa couplings are then given
by
\begin{eqnarray}
\label{yukawa1} \mathcal{L} &\supset
&-\tilde{y}_{q_{ij}^{}}^{}\overline{q_{L_{i}^{}}^{}}
\tilde{\phi}q_{R_{j}^{}}^{}-y_{q_{ij}^{}}^{}\overline{q_{L_{i}^{}}^{}}
\phi q_{R_{j}^{}}^{}
-\tilde{y}_{\psi_{ij}^{}}^{}\overline{\psi_{L_{i}^{}}^{}}
\tilde{\phi}\psi_{R_{j}^{}}^{}
\nonumber\\
&&-y_{\psi_{ij}^{}}^{}\overline{\psi_{L_{i}^{}}^{}}
\phi\psi_{R_{j}^{}}^{}- \frac{1}{2}f_{ij}^{}
\left(\overline{\psi_{L_{i}^{}}^{c}}i\tau_{2}^{}\Delta_{L}^{} \psi_{L_{j}^{}}^{}\right.\nonumber\\
&&\left.+ \overline{\psi_{R_{i}^{}}^{c}}i\tau_{2}^{}\Delta_{R}^{}
\psi_{R_{j}^{}}^{}\right)+\textrm{h.c.}\,,
\end{eqnarray}
where $\tilde{\phi}= \tau_{2}^{}\phi^{\ast}_{}\tau_{2}^{}$,
$y_{q}^{}=y^{\dagger}_{q}$,
$\tilde{y}_{q}^{}=\tilde{y}^{\dagger}_{q}$,
$y_{\psi}^{}=y^{\dagger}_{\psi}$,
$\tilde{y}_{\psi}^{}=\tilde{y}^{\dagger}_{\psi}$ and $f=f^{T}_{}$.
For simplicity, we do not present the most general renormalizable
and parity invariant scalar potential which can be found in many
early works \cite{ms1981,dgko1991}.

\vspace{1mm}

\emph{The seesaw mechanism:} We now review the seesaw mechanism of
the neutrino masses in the left-right symmetric model with the
choice of Higgs scalars we considered. In this case, the left-right
symmetry is broken down to the SM $SU(2)_{L}^{}\times U(1)_{Y}^{}$
symmetry after the right-handed triplet Higgs scalar develops its
vacuum expectation value ($vev$)
$v_{R}^{}\equiv\langle\Delta_{R}^{}\rangle$. From Eq.
(\ref{yukawa1}), we thus have the following Yukawa couplings and
Majorana mass term:
\begin{eqnarray}
\label{yukawa2} \mathcal{L} &\supset
&-y_{d_{ij}^{}}^{}\overline{q_{L_{i}^{}}^{}}
\tilde{\varphi}d_{R_{j}^{}}^{}-y_{u_{ij}^{}}^{}\overline{q_{L_{i}^{}}^{}}
\varphi u_{R_{j}^{}}^{}
-y_{\ell_{ij}^{}}^{}\overline{\psi_{L_{i}^{}}^{}}
\tilde{\varphi}\ell_{R_{j}^{}}^{}
\nonumber\\
&&-y_{\nu_{ij}^{}}^{}\overline{\psi_{L_{i}^{}}^{}}
\varphi\nu_{R_{j}^{}}^{}-\frac{1}{2}f_{ij}^{} v_{R}^{}
\overline{\nu_{R_{i}^{}}^{c}} \nu_{R_{j}^{}}^{}\nonumber\\
&&- \frac{1}{2}f_{ij}^{}
\overline{\psi_{L_{i}^{}}^{c}}i\tau_{2}^{}\Delta_{L}^{}
\psi_{L_{j}^{}}^{}+\textrm{h.c.}\,.
\end{eqnarray}
Here we have defined
\begin{eqnarray}
\phi_{1}^{} = \left[
\begin{array}{c}
\phi^{0}_{1}\\
\phi^{-}_{1}\end{array} \right]\,, ~\phi_{2}^{} = \left[
\begin{array}{c}
\phi^{0\ast}_{2}\\
-\phi^{+\ast}_{2}\end{array} \right]\,,
\end{eqnarray}
and then
\begin{eqnarray}
\label{smhiggs} \varphi&=&\phi_{1}^{}\cos\beta +
\phi_{2}^{}\sin\beta=\left[
\begin{array}{c}
\varphi^{0}_{}\\[1mm]
\varphi^{-}_{}\end{array} \right]\,,\\
\label{fermion1}
y_{d}^{}&=&-y_{q}^{}\sin\beta-\tilde{y}_{q}^{}\cos\beta\,,\\
\label{fermion2}
y_{u}^{}&=&y_{q}^{}\cos\beta+\tilde{y}_{q}^{}\sin\beta\,,\\
\label{fermion3}
y_{\ell}^{}&=&-y_{\psi}^{}\sin\beta-\tilde{y}_{\psi}^{}\cos\beta\,,\\
\label{fermion4}
y_{\nu}^{}&=&y_{\psi}^{}\cos\beta+\tilde{y}_{\psi}^{}\sin\beta\,,
\end{eqnarray}
where
$\displaystyle{\beta=\arctan\left(\frac{v_{2}^{}}{v_{1}^{}}\right)}$
with $v_{1}^{}\equiv\langle\phi_{1}^{}\rangle$ and
$v_{2}^{}\equiv\langle\phi_{2}^{}\rangle$. Obviously, the doublet
scalar $\varphi$ is the SM Higgs. Note that $y_{u}^{}=-y_{d}^{}$ and
$y_{\nu}^{}=-y_{\ell}^{}$ for $\displaystyle{\beta=\frac{\pi}{4}}$
from $v_{1}^{}=v_{2}^{}$. So we will not consider the case of
$v_{1}^{}=v_{2}^{}$ due to the mass differences between the up and
down type quarks.

The first line in Eq. (\ref{yukawa2}) will give the masses to the
charged fermions after the electroweak symmetry is broken by
$v\equiv\langle\varphi\rangle\simeq174\,\textrm{GeV}$. As for the
second line, the first and the second terms generate the Dirac
masses of the neutrinos and the Majorana masses of the right-handed
neutrinos, respectively:
\begin{eqnarray}
\label{diracmass} m_{D}^{}&=& y_{\nu}^{}v\,,\\
\label{rightmass} M_{R}^{}&=& f v_{R}^{}\,.
\end{eqnarray}
For $M_{R}^{}\gg m_{D}^{}$, the left-handed neutrinos can naturally
acquire the small Majorana masses,
\begin{eqnarray}
\label{type1}
m_{\textrm{tree}}^{I}=-m_{D}^{\ast}\frac{1}{M_{R}^{\dagger}}m_{D}^{\dagger}\sim
\mathcal{O}\left(\frac{y_{\nu}^{2}}{f}\right)\frac{v^{2}_{}}{v_{R}^{}}\,,
\end{eqnarray}
i.e. the type-I seesaw formula. The third line will also give the
left-handed neutrinos a Majorana mass term,
\begin{eqnarray}
\label{type2} m_{\textrm{tree}}^{II}= f v_{L}^{} \quad
\textrm{with}\quad v_{L}^{}\equiv\langle\Delta_{L}^{}\rangle\,.
\end{eqnarray}
Here $v_{L}^{}$ can be determined by minimizing the complete scalar
potential \cite{ms1981},
\begin{eqnarray}
\label{leftvev} v_L^{}  \simeq -\frac{\mu\,
v^{2}_{}}{M_{\delta^{0}_{L}}^{2}}\sim -\frac{\mu
v^{2}_{}}{v_{R}^{2}}\,,
\end{eqnarray}
where $\mu\lesssim v_{R}^{}$ is a product of $v_{R}^{}$ and some
combination of couplings entering in the parity invariant scalar
potential. So, we have
\begin{eqnarray}
\label{type2-2} m_{\textrm{tree}}^{II}\sim -f\frac{\mu
v^{2}_{}}{v_{R}^{2}}\,,
\end{eqnarray}
which can be comparable to the type-I seesaw contribution. The
generation of the small $v_{L}^{}$ (\ref{leftvev}) and then the tiny
neutrino masses (\ref{type2-2}) is named as the type-II seesaw.

\vspace{1mm}

\emph{The radiative generation of neutrino masses:} The bi-doublet
Higgs scalar contains two iso-doublet scalars: the SM Higgs
$\varphi$ (\ref{smhiggs}) and
\begin{eqnarray}
\label{doublet} \eta&=&-\phi_{1}^{}\sin\beta +
\phi_{2}^{}\cos\beta=\left[
\begin{array}{c}
\eta^{0}_{}\\
\eta^{-}_{}\end{array} \right]\,.
\end{eqnarray}
$\eta$ couples to the fermions, but it cannot contribute to any
fermion mass at the tree level since it has no $vev$. We shall show
that $\eta$ gives new radiative seesaw contribution to the neutrino
masses through some loop diagrams.

\begin{figure}
\vspace{6.0cm} \epsfig{file=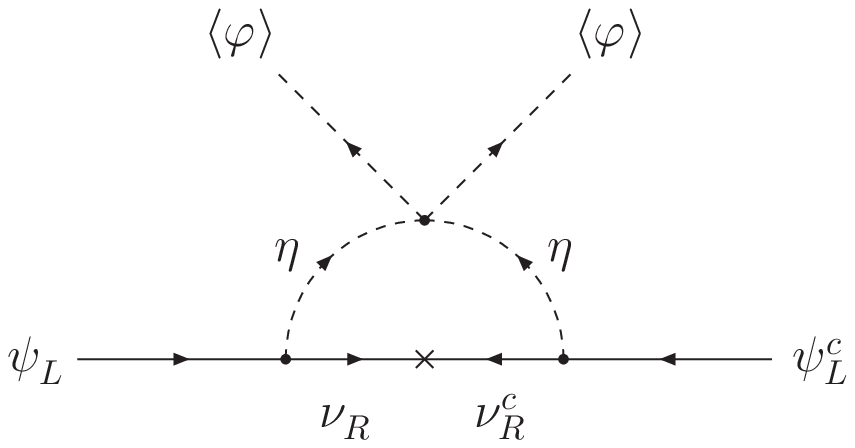, bbllx=5.3cm,
bblly=6.0cm, bburx=15.3cm, bbury=16cm, width=7.0cm, height=7.0cm,
angle=0, clip=0} \vspace{-9.0cm} \caption{\label{massgeneration1}
The one-loop diagram mediated by the right-handed neutrinos for
generating the radiative neutrino masses.}
\end{figure}

For the purpose of demonstration, we deduce the Yukawa couplings of
$\eta$ to the leptons from Eq. (\ref{yukawa1}),
\begin{eqnarray}
\label{yukawa3} \mathcal{L} &\supset&
-h_{\nu_{ij}^{}}^{}\overline{\psi_{L_{i}^{}}^{}}
\eta\nu_{R_{j}^{}}^{}
-h_{\ell_{ij}^{}}^{}\overline{\psi_{L_{i}^{}}^{}}
\tilde{\eta}\ell_{R_{j}^{}}^{}+\textrm{h.c.}\,,
\end{eqnarray}
where
\begin{eqnarray}
\label{fermion5}
h_{\nu}^{}&=&-y_{\psi}^{}\sin\beta+\tilde{y}_{\psi}^{}\cos\beta\nonumber\\
&=&-\left( y_{\ell}^{}\sec 2\beta+\frac{1}{2}y_{\nu}^{}\tan 2\beta\right)~\textrm{for}~\beta\neq\frac{\pi}{4}\,,\\
\label{fermion6}
h_{\ell}^{}&=&-y_{\psi}^{}\cos\beta+\tilde{y}_{\psi}^{}\sin\beta\nonumber\\
&=&-\left(y_{\nu}^{}\sec 2\beta+\frac{1}{2}y_{\ell}^{}\tan
2\beta\right)~\textrm{for}~\beta\neq\frac{\pi}{4}\,.
\end{eqnarray}
As shown in Fig. \ref{massgeneration1}, the quartic coupling between
$\varphi$ and $\eta$,
\begin{eqnarray}
\label{quartic} \mathcal{L}
&\supset&-\lambda(\varphi^{\dagger}_{}\eta)^{2}_{} +\textrm{h.c.}\,,
\end{eqnarray}
where $\lambda \lesssim \mathcal{O}(1)$ is some combination of
couplings entering in the parity invariant scalar potential, will
generate the radiative neutrino masses associated with the first
term in Eq. (\ref{yukawa3}) and the Majorana masses of the
right-handed neutrinos. We choose the basis in which the Majorana
mass matrix (\ref{rightmass}) of the right-handed neutrinos are real
and diagonal and then explicitly calculate the radiative neutrino
masses, which have the same forms with those in the two Higgs
doublet model \cite{ma2006},
\begin{eqnarray}
\label{loop1} \left(m_{\emph{1}-\textrm{loop}}^{I}\right)_{ij}^{}
&=&\sum_{k=1}^{3}\frac{h^{\ast}_{\nu_{ik}^{}}
h^{\ast}_{\nu_{jk}^{}}}{16\pi^{2}_{}}M_{R_{k}^{}}^{}\nonumber\\
&&\times\left[\frac{M_{\eta^{0}_{R}}^{2}}{M_{\eta^{0}_{R}}^{2}-M_{R_{k}^{}}^{2}}
\ln
\left(\frac{M_{\eta^{0}_{R}}^{2}}{M_{R_{k}^{}}^{2}}\right)\right.\nonumber\\
&&\left.-\frac{M_{\eta^{0}_{I}}^{2}}{M_{\eta^{0}_{I}}^{2}-M_{R_{k}^{}}^{2}}
\ln
\left(\frac{M_{\eta^{0}_{I}}^{2}}{M_{R_{k}^{}}^{2}}\right)\right]\,.
\end{eqnarray}
Here $\eta^{0}_{R}$ and $\eta^{0}_{I}$ are defined by
$\displaystyle{\eta^{0}_{}=\frac{1}{\sqrt{2}}\left(\eta^{0}_{R}+i\eta^{0}_{I}\right)}$.
Note the quartic coupling (\ref{quartic}) guarantees the mass
difference between $\eta^{0}_{R}$ and $\eta^{0}_{I}$,
\begin{eqnarray}
\label{difference1}
M_{\eta^{0}_{R}}^{2}-M_{\eta^{0}_{I}}^{2}=4\lambda v^{2}_{}\,,
\end{eqnarray}
where $\lambda$ has been chosen to be real by the proper phase
rotation, so that the radiative neutrino masses (\ref{loop1}) will
not vanish.

The mass splitting between the two doublet scalars $\phi$ and $\eta$
are of the order of $v_{R}^{}$, since they belong to the same
representation of $SU(2)_{R}^{}$. Therefore, $M_{\eta^{0}_{R,I}}^{}$
are of the order of $v_{R}^{}$ for the mass of $\phi$ is much below
$v_R$. In general, $M_{R_{k}^{}}^{}$ is smaller than $v_{R}^{}$, so
we can take $M_{R_{k}^{}}^{2}\ll M_{\eta^{0}_{R,I}}^{2}$ and then
simplify the mass formula (\ref{loop1}) as
\begin{eqnarray}
\label{loop1-2}
\left(m_{\textrm{\emph{1}-loop}}^{I}\right)_{ij}^{}&\simeq&\sum_{k=1}^{3}\frac{h^{\ast}_{\nu_{ik}^{}}
h^{\ast}_{\nu_{jk}^{}}}{16\pi^{2}_{}}M_{R_{k}^{}}^{}\ln
\left(\frac{M_{\eta^{0}_{R}}^{2}}{M_{\eta^{0}_{I}}^{2}}\right)\nonumber\\
&\simeq&\sum_{k=1}^{3}\frac{h^{\ast}_{\nu_{ik}^{}}
h^{\ast}_{\nu_{jk}^{}}}{16\pi^{2}_{}}M_{R_{k}^{}}^{}\ln
\left[1+\mathcal{O}\left(\lambda\frac{v^{2}_{}}{v_{R}^{2}}\right)\right]\nonumber\\
&\sim&\frac{1}{16\pi^{2}_{}}\mathcal{O}\left(\lambda
h_{\nu}^{2}f\right)\frac{v^{2}_{}}{v_{R}^{}}\,.
\end{eqnarray}

\begin{figure}
\vspace{7.0cm} \epsfig{file=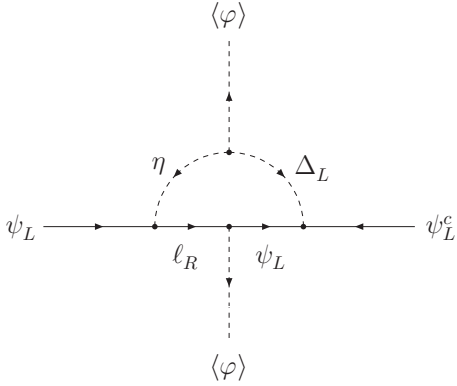, bbllx=5.3cm,
bblly=6.0cm, bburx=15.3cm, bbury=16cm, width=7.0cm, height=7.0cm,
angle=0, clip=0} \vspace{-7.5cm} \caption{\label{massgeneration2}
The one-loop diagram mediated by the left-handed triplet Higgs for
generating the radiative neutrino masses.}
\end{figure}

Now we discuss the contribution from the left-handed triplet Higgs
to the radiative neutrino masses. There is a cubic coupling among
$\Delta_{L}^{}$, $\varphi$ and $\eta$,
\begin{eqnarray}
\label{cubic1} \mathcal{L}
&\supset&-\mu'\eta^{T}_{}i\tau_{2}^{}\Delta_{L}^{}\varphi+\textrm{h.c.}\,,
\end{eqnarray}
where $\mu'\lesssim v_{R}^{}$ is a product of $v_{R}^{}$ and some
combination of couplings entering in the parity invariant scalar
potential. Therefore, associated with the third term in Eq.
(\ref{yukawa2}) and the second term in Eq. (\ref{yukawa3}), the
cubic coupling (\ref{cubic1}) can generate the radiative neutrino
masses as shown in Fig. \ref{massgeneration2}.

For illustration, we write down the mass matrix of $\delta_{L}^{+}$
and $\eta^{-}_{}$,
\begin{eqnarray}
\mathcal{L}&\supset&-\left[
 \delta^{+\ast}_{L}\,,~\eta^{-}_{}\right]\left[
\begin{array}{cc}
 M_{\delta^{+}_{L}}^{2}& -\frac{1}{\sqrt{2}}\mu'v\\
 -\frac{1}{\sqrt{2}}\mu'v & M_{\eta^{-}_{}}^{2}\end{array}
\right]\left[\begin{array}{c}
 \delta_{L}^{+} \\
 \eta^{-\ast}_{}
\end{array}\right]\,,
\end{eqnarray}
Here $\mu'$ has been chosen to be real by the proper phase rotation.
There are two mass eigenstates $S_{1}^{}$ and $S_{2}^{}$,
\begin{eqnarray}
\left(\begin{array}{c}
 S_{1}^{} \\
 S_{2}^{} \end{array}\right)&=& \left(
\begin{array}{cc}
 \cos\vartheta & -\sin\vartheta\\
 \sin\vartheta & \cos\vartheta\end{array}
\right)\left(\begin{array}{c}
 \delta^{+}_{L} \\
 \eta^{-\ast}_{} \end{array}\right)\,,
\end{eqnarray}
with the mixing angle
\begin{eqnarray}
\label{mixing1} \tan2\vartheta=\frac{\sqrt{2}\mu'
v}{M_{\delta^{+}_{L}}^{2}-M_{\eta^{-}_{}}^{2}}\,,
\end{eqnarray}
and the masses
\begin{eqnarray}
M_{S_{1,2}}^{2}&=&\frac{1}{2}\left[M_{\delta^{+}_{L}}^{2}+M_{\eta^{-}_{}}^{2}\right.\nonumber\\
&&\left.\pm\sqrt{\left(M_{\delta^{+}_{L}}^{2}-M_{\eta^{-}_{}}^{2}\right)^{2}_{}
+2\mu'^{2}_{}v^{2}_{}}\right]\,.
\end{eqnarray}
For $ M_{\delta^{+}_{L}}^{2}\sim M_{\eta^{-}_{}}^{2}\sim
M_{\delta^{+}_{L}}^{2}-M_{\eta^{-}_{}}^{2}=\mathcal{O}(v_{R}^{2})$
and $\mu'v\ll v_{R}^{2}$, we have
\begin{eqnarray}
\label{mixing2} &&\vartheta=
\mathcal{O}(\frac{\mu'v}{v_{R}^{2}})\,,\\
\label{difference2} &&M_{S^{}_{1,2}}^{2}\sim M_{S^{}_{2}}^{2}-
M_{S^{}_{1}}^{2}=\mathcal{O}(v_{R}^{2})\,.
\end{eqnarray}

We then calculate the formula of the radiative neutrino masses
induced by Fig. \ref{massgeneration2},
\begin{eqnarray}
\label{loop2} \left(m_{\emph{1}-\textrm{loop}}^{II}\right)_{ij}^{}
&=&\frac{1}{16\pi^2_{}}\frac{\sin2\vartheta}{\sqrt{2}}
\sum_{k=e,\mu,\tau}^{}f_{ik}^{}h^{\dagger}_{\ell_{kj}^{}}
m_{k}^{}\nonumber\\
&&\times\left[\frac{M_{S_{1}^{}}^{2}}{M_{S_{1}^{}}^{2}-m_{k}^{2}}
\ln
\left(\frac{M_{S_{1}^{}}^{2}}{m_{k}^{2}}\right)\right.\nonumber\\
&&\left. - \frac{M_{S_{2}^{}}^{2}}{M_{S_{2}^{}}^{2}-m_{k}^{2}} \ln
\left(\frac{M_{S_{2}^{}}^{2}}{m_{k}^{2}}\right)\right]\,.
\end{eqnarray}
Here we have chosen the basis in which the Yukawa couplings
(\ref{fermion3}) of the charged leptons are real and diagonal and
have referred $m_{k}^{}$ to the masses of the charged leptons. Note
the above mass formula is different from that of the Zee model
\cite{zee1980} becaue $f$ is symmetric. For $M_{S_{1,2}^{}}^{2}\gg
m_{k}^{2}$, we simplify Eq. (\ref{loop2}) as
\begin{eqnarray}
\label{loop2-2}
&&\left(m_{\emph{1}-\textrm{loop}}^{II}\right)_{ij}^{}\nonumber\\
&\simeq&\frac{1}{16\pi^2_{}}\frac{\sin2\vartheta}{\sqrt{2}}
\sum_{k=e,\mu\,\tau}^{}f_{ik}^{}h^{\dagger}_{\ell_{kj}^{}} m_{k}^{}
\ln
\left(\frac{M_{S_{2}^{}}^{2}}{M_{S_{1}^{}}^{2}}\right)\nonumber\\
&\simeq&\frac{1}{16\pi^2_{}}\mathcal{O}\left(\frac{\mu'v}{v_{R}^{2}}\right)
\ln\left[1+\mathcal{O}\left(1\right)\right]\sum_{k=e,\mu,\tau}^{}
f_{ik}^{}h^{\dagger}_{\ell_{kj}} m_{k}^{}\nonumber\\
&\sim&\frac{1}{16\pi^2_{}}\frac{\mu'v^{2}_{}}{v_{R}^{2}}\mathcal{O}
\left(y_{\ell}^{}h_{\ell}^{}f\right)
\end{eqnarray}
by using Eqs. (\ref{mixing2}) and (\ref{difference2}).

\begin{figure}
\vspace{5.0cm} \epsfig{file=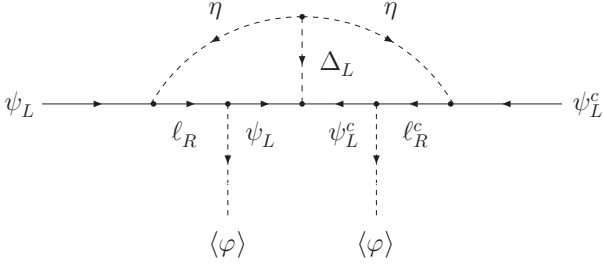, bbllx=5.3cm,
bblly=6.0cm, bburx=15.3cm, bbury=16cm, width=7.0cm, height=7.0cm,
angle=0, clip=0} \vspace{-7.5cm} \caption{\label{massgeneration3}
The two-loop diagram for generating the radiative neutrino masses.}
\end{figure}

In addition to the two one-loop diagrams, i.e. Figs.
\ref{massgeneration1} and \ref{massgeneration2}, there is a two-loop
diagram as shown in Fig. \ref{massgeneration3} contributing to the
radiative neutrino masses due to the following cubic coupling,
\begin{eqnarray}
\label{cubic2} \mathcal{L}
&\supset&-\mu''\eta^{T}_{}i\tau_{2}^{}\Delta_{L}^{}\eta+\textrm{h.c.}\,,
\end{eqnarray}
where $\mu''\lesssim v_{R}^{}$ is a product of $v_{R}^{}$ and some
combination of couplings entering in the parity invariant scalar
potential. For simplicity, we choose $\mu''$ to be real by the
proper phase rotation. Similar to the Zee-Babu model \cite{zee1985},
we calculate the two-loop induced neutrino masses as below,
\begin{eqnarray}
\label{loop3}
&&\left(m_{\emph{2}-\textrm{loop}}^{}\right)_{ij}^{}\nonumber\\
&=&\frac{1}{64\pi^{4}_{}}\sum_{k,n=e,\mu,\tau}^{}
h_{\ell_{ik}^{}}^{\ast}f_{kn}^{}h_{\ell_{nj}^{}}^{\dagger}\frac{\mu''m_{k}^{}m_{n}^{}}{M_{\delta^{++}_{L}}^{2}}\nonumber\\
&&\times\left\{\sin^{4}_{}\vartheta\,
\left[\ln\left(1+\frac{M_{\delta^{++}_{L}}^{2}}{M_{S_{1}^{}}^{2}}\right)\right]^{2}_{}\right.\nonumber\\
&&+\frac{1}{2}\sin^{2}_{}2\vartheta\,
\ln\left(1+\frac{M_{\delta^{++}_{L}}^{2}}{M_{S_{1}^{}}^{2}}\right)
\ln\left(1+\frac{M_{\delta^{++}_{L}}^{2}}{M_{S_{2}^{}}^{2}}\right)
\nonumber\\
&&\left.+\cos^{4}_{}\vartheta\,
\left[\ln\left(1+\frac{M_{\delta^{++}_{L}}^{2}}{M_{S_{2}^{}}^{2}}\right)\right]^{2}_{}
\right\}\nonumber\\
&\simeq&\frac{1}{64\pi^{4}_{}}\sum_{k,n=e,\mu,\tau}^{}
h_{\ell_{ik}^{}}^{\ast}f_{kn}^{}h_{\ell_{nj}^{}}^{\dagger}\frac{\mu''m_{k}^{}m_{n}^{}}{M_{\delta^{++}_{L}}^{2}}\mathcal{O}\left(1\right)\nonumber\\
&\sim&
\frac{1}{64\pi^{4}_{}}\frac{\mu''v^{2}_{}}{v_{R}^{2}}\mathcal{O}(y_{\ell}^{2}h_{\ell}^{2}f)\,.
\end{eqnarray}
Here we have taken $M_{\delta^{++}_{L}}^{2}\sim
M_{S_{1,2}^{}}^{2}\sim v_{R}^{2}\gg m_{e,\mu,\tau}^{2}$ and
$\vartheta\ll 1$ into account. Unlike the Zee-Babu model
\cite{zee1985}, we needn't constrain the Yukawa couplings
$h_{\ell}^{}$ to be asymmetric.

\vspace{1mm}

\emph{The radiative neutrino masses versus the tree-level neutrino
masses:} Now the complete neutrino masses should include five parts,
\begin{eqnarray}
\label{completemass}
m_{\nu}^{}&=&m_{\textrm{tree}}^{I}+m_{\textrm{tree}}^{II}+m_{\emph{1}-\textrm{loop}}^{I}
+m_{\emph{1}-\textrm{loop}}^{II}+m_{\emph{2}-\textrm{loop}}^{}\nonumber\\
&\sim&\frac{v^{2}_{}}{v_{R}^{}}\left[\mathcal{O}\left(\frac{y^{2}_{\nu}}{f}\right)
+\frac{\mu}{v_{R}^{}}\mathcal{O}\left(f\right)+\frac{1}{16\pi^{2}_{}}\mathcal{O}\left(\lambda
h_{\nu}^{2}f\right)\right.\nonumber\\
&&
\left.+\frac{1}{16\pi^{2}_{}}\frac{\mu'}{v_{R}^{}}\mathcal{O}\left(y_{\ell}^{}h_{\ell}^{}f
\right)+\frac{1}{64\pi^{4}_{}}\frac{\mu''}{v_{R}^{}}\mathcal{O}(y_{\ell}^{2}h_{\ell}^{2}f)\right]\,,
\end{eqnarray}
where $v\simeq 174\,\textrm{GeV}$,
$v_{R}^{}>\mathcal{O}(\textrm{TeV})$ and
$y_{\ell}^{}\lesssim\mathcal{O}(10^{-2}_{})$ have been known. In the
following, we shall show that the five parts can have different
weight depending on the choice of the parameters. In particular, for
some choice, the radiative contributions could dominate over the
tree-level seesaw and could explain the observed neutrino masses.

We now demonstrate that the loop-induced neutrino masses could
dominate over the tree-level seesaw for some choice of the
parameters. For naturalness, we further assume that there is no
cancelation in Eqs. (\ref{fermion1}) and (\ref{fermion2}) to
generate a quark mass hierarchy so that $v_{1}^{}$ and $v_{2}^{}$
should not be at the same order since the top quark is much heavier
than the bottom quark. For example, we will take
$\displaystyle{\frac{v_{2}^{}}{v_{1}^{}}=\mathcal{O}(10^{-2}_{})}$
and hence $h_{\nu}^{}\sim y_{\ell}^{}+10^{-2}_{}y_{\nu}^{}$ and
$h_{\ell}^{}\sim y_{\nu}^{}+10^{-2}_{}y_{\ell}^{}$ in the
quantitative estimation. We then find: (a)
$m_{\emph{1}-\textrm{loop}}^{I}\gtrsim m_{\textrm{tree}}^{I} $ for
$f=\mathcal{O}(0.1)$, $\lambda\lesssim \mathcal{O}(1)$ and
$y_{\nu}^{}\lesssim \mathcal{O}(10^{-4}_{})$; (b)
$m_{\emph{1}-\textrm{loop}}^{II}\gtrsim m_{\textrm{tree}}^{I}$ for
$f=\mathcal{O}(0.1)$, $\mu'\lesssim v_{R}^{}$ and
$y_{\nu}^{}\lesssim \mathcal{O}(10^{-5}_{})$; (c)
$m_{\emph{2}-\textrm{loop}}^{}\gtrsim m_{\textrm{tree}}^{I} $ for
$f=\mathcal{O}(0.1)$, $\mu''\lesssim v_{R}^{}$ and
$y_{\nu}^{}\lesssim \mathcal{O}(10^{-9}_{})$. Obviously, we can have
$m_{\emph{1}-\textrm{loop}}^{I,II}\gtrsim m_{\textrm{tree}}^{II}$
and $m_{\emph{2}-\textrm{loop}}^{}\gtrsim m_{\textrm{tree}}^{II}$
for the proper $\mu$ and other parameters.

We then choose some values of the unknown parameters to show that
the loop contributions can match the observed neutrino masses: (i) $
m_{\emph{1}-\textrm{loop}}^{I}\sim
\mathcal{O}(10^{-2}_{}\,\textrm{eV})\gg m_{\textrm{tree}}^{I}\sim
m_{\textrm{tree}}^{II}\sim m_{\emph{1}-\textrm{loop}}^{II}\gg
m_{\emph{2}-\textrm{loop}}^{}$ for
$v_{R}^{}=\mathcal{O}(10^{8}_{}\,\textrm{GeV})$,
$\mu=\mathcal{O}(\textrm{GeV})$, $\mu'\lesssim v_{R}^{}$,
$\mu''\lesssim v_{R}^{}$, $\lambda=\mathcal{O}(1)$,
$f=\mathcal{O}(0.1)$ and $y_{\nu}^{}=\mathcal{O}(10^{-5}_{})$; (ii)
$m_{\textrm{tree}}^{I}\sim m_{\textrm{tree}}^{II}\sim
m_{\emph{1}-\textrm{loop}}^{I}\sim
m_{\emph{1}-\textrm{loop}}^{II}\sim
\mathcal{O}(10^{-2}_{}\,\textrm{eV}) \gg
m_{\emph{2}-\textrm{loop}}^{}$ for
$v_{R}^{}=\mathcal{O}(10^{4}_{}\,\textrm{GeV})$,
$\mu=\mathcal{O}(10^{-6}_{}\,\textrm{GeV})$,
$\mu'=\mathcal{O}(10^{2}_{}\,\textrm{GeV})$, $\mu''\lesssim
v_{R}^{}$, $\lambda=\mathcal{O}(10^{-4}_{})$, $f=\mathcal{O}(0.1)$
and $y_{\nu}^{}=\mathcal{O}(10^{-6}_{})$; (iii)
$m_{\emph{1}-\textrm{loop}}^{I}\sim
m_{\emph{1}-\textrm{loop}}^{II}\sim
m_{\emph{2}-\textrm{loop}}^{}\sim
\mathcal{O}(10^{-2}_{}\,\textrm{eV})\ll m_{\textrm{tree}}^{I}\sim
m_{\textrm{tree}}^{II}$ for
$v_{R}^{}=\mathcal{O}(10^{4}_{}\,\textrm{GeV})$, $\mu\sim v_{R}^{}$,
$\mu'=\mathcal{O}(0.1\,\textrm{GeV})$, $\mu''\sim v_{R}^{}$,
$\lambda=\mathcal{O}(10^{-4}_{})$, $f=\mathcal{O}(0.1)$ and
$y_{\nu}^{}=\mathcal{O}(0.1)$. In the last case, we need the
fine-tuned cancelation of $m_{\textrm{tree}}^{I}$ and
$m_{\textrm{tree}}^{II}$ to ensure the complete neutrino masses
below the experimental limit.

\vspace{1mm}

\emph{Summary:} We find the new radiative seesaw mechanism for the
neutrino masses in the conventional left-right symmetric model with
one bi-doublet, one left-handed triplet and one right-handed triplet
Higgs scalars. Specifically the neutrino masses can be generated not
only by the tree-level seesaw but also by two one-loop diagrams and
one two-loop diagram. For some choice of the parameters, the
observed neutrino masses can be explained by these loop
contributions.

\textbf{Acknowledgments}: We thank Goran Senjanovi$\rm\acute{c}$ for
helpful discussions.

\end{document}